\documentclass[sigconf]{acmart}
\usepackage{hyperref} 
\usepackage{hyperxmp}
\usepackage{longtable}
\usepackage{multirow}

\AtBeginDocument{%
  \providecommand\BibTeX{{%
    \normalfont B\kern-0.5em{\scshape i\kern-0.25em b}\kern-0.8em\TeX}}}

\setcopyright{acmcopyright}
\copyrightyear{2025}
\acmYear{2025}

\acmConference[CHI '25]{CHI Conference on Human Factors in Computing Systems}{April 26-May 1, 2025}{Yokohama, Japan}
\acmBooktitle{CHI Conference on Human Factors in Computing Systems (CHI '25), April 26-May 1, 2025, Yokohama, Japan}\acmDOI{10.1145/3706598.3713726}
\acmISBN{979-8-4007-1394-1/25/04}

\begin{document}
\title{Large Language Models in Qualitative Research: \\Uses, Tensions, and Intentions}
\author{Hope Schroeder}
\authornote{Both authors contributed equally to this research.}
\email{hopes@mit.edu}
\affiliation{%
  \institution{MIT}
  \city{Cambridge}
  \state{Massachusetts}
  \country{USA}}

\author{Marianne Aubin Le Quéré}
\email{msa258@cornell.edu}
\authornotemark[1]
\affiliation{%
    \institution{Cornell Tech}
    \city{New York City}
    \state{New York}
    \country{USA}}

\author{Casey Randazzo}
\email{cer124@rutgers.edu}
\affiliation{%
\institution{Rutgers University}
\city{New Brunswick}
\state{New Jersey}
\country{USA}}

\author{David Mimno}
\email{mimno@cornell.edu}
\affiliation{%
\institution{Cornell University}
\city{Ithaca}
\state{New York}
\country{USA}}

\author{Sarita Schoenebeck}
\email{yardi@umich.edu}
\affiliation{%
\institution{University of Michigan}
\city{Ann Arbor}
\state{Michigan}
\country{USA}}

\renewcommand{\shortauthors}{Schroeder and Aubin Le Qu\'er\'e, et al.}

\begin{abstract}

Qualitative researchers use tools to collect, sort, and analyze their data. Should qualitative researchers use large language models (LLMs) as part of their practice? LLMs could augment qualitative research, but it is unclear if their use is appropriate, ethical, or aligned with qualitative researchers' goals and values. We interviewed twenty qualitative researchers to investigate these tensions. Many participants see LLMs as promising interlocutors with attractive use cases across the stages of research, but wrestle with their performance and appropriateness. 
Participants surface concerns regarding the use of LLMs while protecting participant interests, and call attention to an urgent lack of norms and tooling to guide the ethical use of LLMs in research. 
We document the rapid and broad adoption of LLMs across surfaces, which can interfere with intentional use vital to qualitative research.
We use the tensions surfaced by our participants to outline recommendations for researchers considering using LLMs in qualitative research and design principles for LLM-assisted qualitative research tools.
\end{abstract}



\maketitle

\section{Introduction}

The qualitative research process is grounded in deep engagement with participants and their data through an iterative sensemaking process.
Today, Large Language Models (LLMs), particularly those accessed through chat-based interfaces like ChatGPT~ \cite{chatgpt}, are popular tools in research-related tasks, including within the social sciences \cite{grossmann2023ai, liao2024llms}.
While a growing number of dedicated tools for qualitative work use LLMs to streamline research processes~\cite{gebreegziabher2023patat, gao2024collabcoder, choksi2024under}, direct interactions with ChatGPT and similar interfaces may also be increasingly used by qualitative researchers, a topic which has not received thorough investigation to this point.

LLMs distinguish themselves from prior tools for qualitative research.
First, chat-based LLMs are marketed as general purpose tools, not dedicated qualitative analysis tools. Users interact with them in natural language, unlike most dedicated software platforms. LLMs go beyond the simple document-organization or analysis functions in classic tools, and facilitate highly flexible interpretations of documents, such as the ability to produce summaries or annotations, behaviors which mimic aspects of the human sensemaking process.
Second, LLM use is currently dominated by one product, ChatGPT, that has rapidly entered general use by the public~\cite{Porter2023}, including by qualitative researchers.
Third, the dominant user interface metaphor~\cite{neale1997role} of ChatGPT and its competitors is that of a jack-of-all-trades assistant and/or collaborator. This pattern is intuitive thanks to the conversational interface, yet completely unguided. 
% LLMs are not simply about speed, efficiency, processing, or other typical functions of software tools. 
With wide-ranging and open-ended interactions with LLMs increasingly permeating everyday life, qualitative researchers are left to wonder whether or how these tools might apply to their research process or outputs. 
While some qualitative researchers support integrating LLMs into aspects of qualitative research, others, as ~\citet{soden2024evaluating} highlight, will be ``concerned that increasingly all qualitative research, including that which draws on interpretive traditions of research design, is being evaluated from the perspective of positivism.''

There is a critical and immediate need to understand the degree to which LLMs are shaping qualitative research processes.
On the one hand, many researchers may benefit from the inclusion of LLM-based features in qualitative data analysis (QDA) systems, as has been explored in HCI literature~\cite{gao2024collabcoder, jiang2021supporting, overney2024sensemate}.
However, the recent explosion of work using LLMs to annotate text in more complex ways than previously possible~\cite{ziems2024can, bail_can_2024} may risk the excitement of communities of scholars and tool builders without full consideration for potential tradeoffs to the sensemaking process.
In particular, readily available tools that seek to automate the qualitative sensemaking process could decrease the amount of time researchers spend with their participant data, a potential risk to qualitative researchers' depth of understanding.

Furthermore, the speed of LLM development has outpaced guidance on their ethical use~\cite{baronchelli2024shaping}.
Most policies currently guiding use focus on the danger of uploading sensitive information to cloud services~\cite{ollion2024dangers} or text generation in publications~\cite{hosseini2023fighting}.
The HCI community is contending with the need for developed policies that tackle how to use AI ethically in research~\cite{shen2023shaping, aubin2024llms}.
% While many have raised concerns about the ethics of these tools, few have presented
While guidelines are beginning to emerge, there are no guidelines specific to qualitative work today.

This work critically reflects on the role of LLMs in qualitative research and offers considerations and recommendations for researchers considering using LLMs.
Through our work, we answer the following research questions:
\begin{itemize}
    \item RQ1: In what ways are LLMs being adopted into qualitative research processes today, and how might they be incorporated in the future?
    \item RQ2: What do qualitative researchers say they stand to lose or gain from adopting LLMs into their research processes, and what tensions might arise?
    \item RQ3: What are potential models for the ethical incorporation of LLMs into the qualitative research process?
\end{itemize}

Qualitative research is a broad and rich tradition with scholars from varied disciplines and ways of thinking. This paper focuses on the subset of researchers, including mixed-method researchers, who are considering using LLMs in their qualitative research processes. Those researchers should be clear-eyed about what is gained and lost when LLMs are used in data collection, analysis, or writing. 

This work surfaces both the curiosities and concerns of qualitative and mixed-method researchers across disciplines who are wrestling with what the introduction of LLMs means for their qualitative research process.
Through twenty semi-structured interviews with researchers across academic communities, we sought to understand any current use, motivations for use, perceptions of usefulness, and concerns about the use of LLMs in qualitative research.
Our findings show that qualitative researchers can envision leveraging LLMs for their work to help with a range of uses including generating interview materials and analyzing research artifacts. 
Specifically, the interactivity of ChatGPT has spurred researchers to think creatively about what chat-based interfaces can bring to their work.
Simultaneously, researchers harbor fundamental concerns about ethics, unequal adoption of new technologies, model bias, and performance, leading some to abstain from the use of LLMs to conduct qualitative research, and others to experiment despite concerns regarding participant privacy.
% We conclude that we are witnessing a moment of structural shift, where new tools and practices are being used, but scholarly communities and their dedicated tools are not equipped for conducting ethical and effective research with them.
New tools and practices are rapidly being adopted, but scholarly communities are not yet equipped to conduct ethical and effective research with them. As such, we are at a junction that urgently requires thorough evaluation and recalibration of both our methodological approaches and our thinking on research practices.
As a step forward, we outline considerations and recommendations for future qualitative researchers interested in using LLMs, and we present design opportunities and principles for creating LLM-based qualitative data analysis tools given our findings.

\section{Related Work}

% \subsection{Qualitative Research and New Technologies in HCI}

Qualitative research is a broad term with different definitions and applications across disciplines~\cite{hammersleyWhatQualitativeResearch2013}. 
In the late 1960s, the method emerged in retaliation to quantitative and positivist approaches to science characterized by hypothesis testing and objectivity~\cite{hammersleyRelationshipQualitativeQuantitative1996}. 
HCI scholars often turn to qualitative methods when studying populations or contexts that require deep exploration and understanding of experiences, behaviors, and interactions \cite{randazzo_if_2023, randazzo_et_al_2023, rankinExploringPluralityBlack2019, freemanStreamingYourIdentity2020, bardzellFeministHCIMethodology2011, bardzellPleasureYourBirthright2011}. 

In contrast to quantitative methods that prioritize numerical data and generalizability, qualitative research aims to uncover how participants perceive, understand, and make meaning of their experiences \cite{jaccard_theory_2020}.
Qualitative researchers often utilize diverse techniques, such as interviews, observations, and narratives \cite{alvarez2023binary, nolan2024mal} to capture the complexity of lived experiences \cite{braun2013successful}. 
The epistemological foundation of qualitative research is \textit{social constructionism} \cite{berger1966reality}, in which meaning and knowledge is subjective and constructed in and through communication \cite{weinberg2009social}, enabling researchers to capture the social contexts that shape participants’ lives \cite{tracy2024qualitative}. 
Through deep engagement with participants’ worlds, qualitative scholars seek to describe and interpret phenomena in a way that reflects participants’ understanding of their experiences \cite{jaccard_theory_2020}, while also contributing to theory development through inductive analysis \cite{scharp2021thematic}. 

As a field, HCI is both home to researchers with qualitative approaches as well as mixed-method researchers, a fact which has sometimes led to tension in the community regarding how contributions should be evaluated.
% allowed for novel studies but also kept tensions aflame regarding 
In particular, a stigma surrounding the perceived rigor of qualitative research \cite{charmazPursuitQualityGrounded2021} has sparked debates over how best to evaluate such methods \cite{kvaleSocialConstructionValidity1995, mcdonald_reliability_2019, rolfeValidityTrustworthinessRigour2006}. 
As a result, conducting qualitative research can involve a complex and sometimes creative negotiation with the norms and evaluative criteria of one's research communities \cite{denzinSAGEHandbookQualitative2024}. 
Within the HCI community, there have been concerns that increasingly quantitative evaluation techniques, such as providing participant demographics, are being unduly required of qualitative scholars~\cite{soden2024evaluating}.
LLMs may lend the impression that qualitative inquiry can be automated, and their integration into Qualitative Data Analysis (QDA) software may increasingly impose positivist approaches that conflict with interpretivist traditions.

% \subsection{Technology Practices and Tools in Qualitative Research} 
Approaches to analyzing qualitative data (e.g., grounded theory, thematic analysis) were originally designed for manual, pen-and-paper applications \cite{deterdingFlexibleCodingIndepth2021}.
In the long history of qualitative research practice, many new technologies have been introduced, including the recording and digitization of interview data, popularization of automated transcription, and introduction of phone interviews, eventually followed by online interviews~\cite{brown2002going, bryda2023qualitative}.
The widespread adoption of software tools for qualitative data analysis marked a significant and, at times, contentious shift in the field \cite{gilbertToolsAnalyzingQualitative2014}. 
Early critiques of qualitative software centered around concerns that these tools created too much distance from the data~\cite{gilbert2002going}, compromised the depth of analysis~\cite{blismas2003computer}, or could be misused by researchers who lacked an understanding of them~\cite{gilbertToolsAnalyzingQualitative2014}.

Advancements in natural language processing (NLP) and machine learning have widened methods for data analysis over the past several decades, and the sudden rise of LLMs has, once again, opened up new possibilities. 
LLMs are being used for text annotation in interpretive contexts~\cite{ziems2024can, bail_can_2024}, including qualitative coding~\cite{choksi2024under, dunivin2024scalable}.
Past work has considered both the potential tensions and compatibilities between computational tools and interpretive work, grounding the conversation regarding the potential usefulness of computational tools in ways that can support qualitative work~\cite{baumer_comparing_2017, nelson2020computational}.
In HCI, tools that leverage advancements in machine learning have been developed to assist qualitative researchers with data and text analysis since before the recent LLM era.
The new generation of LLMs has ushered in new kinds of tools, with developers creating tools for qualitative researchers that bear researcher goals in mind, like agency over the coding process~\cite{overney2024sensemate},  convergent discussion on coding among collaborators~\cite{gao2024collabcoder}, or incorporating trauma-informed design principles~\cite{tseng2025mitigating}. 
Studies introducing and evaluating these tools contain insights regarding the promises and pitfalls of LLMs in qualitative research, but largely focus on users' perceptions of the particular tools being developed for a task-specific use case, such as data annotation.
Given the sudden, widespread ubiquity of LLMs through task-agnostic interfaces like ChatGPT, experiences with AI tools have become more common while being less scaffolded by purpose-built software, warranting investigation of current LLM use by qualitative researchers in a task-agnostic context.

Beyond the evaluation of specific software interfaces, prior work has also explored the concept of using AI tools to augment qualitative work.
\citet{jiang2021supporting} and \citet{tools_in_place} tackle the question of how qualitative researchers may respond to incorporating AI in their workflows.
\citet{jiang2021supporting} identify that AI tools for qualitative work should ``honor serendipity, human agency, and ambiguity.'' \citet{tools_in_place} highlight that intentional use and ordering of AI tools in qualitative tasks is paramount, and that the use of any computational tools may entail shifts in qualitative processes towards scale, abstraction, and delegation.
% that intentional use of AI tools in qualitative contexts can significantly diverge between research approaches and traditions.
These works lay key groundwork for this study, but were both published prior to the popular adoption of LLMs.
% MALQ: PRESUPPOSE DEDICATED INTERFACES, NO SCAFFOLDING OF TASKS!
Both papers presuppose the use of AI in the qualitative research workflow in bounded, task-specific ways, an idea which largely contrasts with the fluid ways that researchers often use LLMs throughout workflows today.
% Given the sudden, widespread ubiquity of LLMs through task-agnostic interfaces like ChatGPT, experiences with AI tools have become more common for all kinds of users, including those not in computing. 
The sudden adoption of LLMs warrants an urgent re-examination of AI tools in qualitative work, with early research suggesting that 81\% of researchers use LLMs in their research practices~\cite{liao2024llms}. Many papers have examined the application of an LLM, or of ChatGPT in particular, to the problem of a thematic analysis for an applied case, with examples across fields as varied as medicine and education~\cite{PENTUCCI2024HYB}. Researchers often report promise regarding LLMs' ability to capture some or most themes humans do, but also perceive a lack of depth, nuance, or variety in identified themes \cite{spangler2024analyzing, morgan, jalali2024integrating, li2024comparing, PENTUCCI2024HYB, wachinger2024prompts}. Simultaneously, some work raises or notes researchers' concerns regarding the reliability and validity of using LLMs in qualitative research, and privacy concerns that may arise when using them \cite{chatgpt_thematic, friedman2024should}.
As LLMs permeate society, sociotechnical gaps~\cite{ackerman2000intellectual} may arise from the use of popular tools like ChatGPT for qualitative research purposes, where the social expectations of research do not match the underlying technical capabilities~\cite{liao2023rethinking, ehsan2023charting}.
In light of these societal shifts, we endeavor a fresh examination of the current uses, attitudes, and tensions that arise when using LLMs in qualitative research.
Through highlighting a range of considerations for qualitative researchers to scaffold their intentional use of these tools, we chart a path forward for using LLMs conscientiously in qualitative work.

\section{Methods}

To understand how qualitative researchers are interacting with generative AI tools or choosing not to, we conducted a series of twenty semi-structured interviews in June-August 2024.
We analyzed the interviews using inductive thematic analysis, inspired by grounded theory ~\cite{braun2006using}, in which concepts emerge from the data \cite{jaccard_theory_2020}.
These interview procedures were approved by the Institutional Review Board at Cornell University.

Participants were recruited through personal networks, email outreach about past work, and through in-person events and conferences.
The inclusion criterion for this interview was simple: we sought to speak with researchers who consider themselves qualitative scholars, including mixed-method researchers.
% \del{researchers who use qualitative methods as part of a mixed methods approach to their work.}
Since we wanted to include a variety of perspectives, we did not recruit participants based on their use or knowledge of LLMs, and we recruited participants across academic domains.
Since qualitative research perspectives can vary by field, we considered it important to allow potential interviewees to self-define as qualitative researchers.

These recruitment strategies resulted in twenty participant interviews, summarized in Table~\ref{tab:participants}.
Our participants spanned seven academic fields, include sixteen PhD students, and are balanced between nine primarily qualitative and eleven mixed-method scholars.
% Although we did not 
Beyond this high-level information, we did not explicitly collect demographic information from participants.
% Our primary aim in recruitment was to balance between different fields.
Nonetheless, we note that a majority of participants presented as women and as researchers of color, and five participants referenced being non-native English speakers in interviews.
Outreach in our personal networks resulted in a population that was largely located in North American Universities, with one exception.

\begin{footnotesize}
\begin{table*}[]
\begin{tabular}{@{}llll@{}}
\toprule
Participant ID & Academic Domain & Position & Approach to Research \\ \midrule
P1 & HCI & PhD Student & Mixed-Method \\
P2 & Communication & PhD Student & Mixed-Method \\
P3 & Political Science & Research project manager & Primarily Qualitative \\
P4 & HCI & PhD Student & Mixed-Method   \\
P5 & Social Work & PhD Student & Mixed-Method  \\ 
P6 & Communication & PhD Student & Primarily Qualitative  \\ 
P7 & Political Science & PhD Student & Mixed-Method \\
P8 & HCI & PhD Student & Primarily Qualitative \\
P9 & HCI & Professor & Mixed-Method  \\
P10 & HCI & PhD Student & Primarily Qualitative  \\
P11 & HCI & PhD Student & Primarily Qualitative  \\
P12 & HCI & PhD Student & Primarily Qualitative  \\
P13 & Sociology & Research Scholar & Mixed-Method  \\
P14 & Political Science & PhD Student & Mixed-Method \\
P15 & Sociology & Former Professor & Primarily Qualitative  \\
P16 & Communication & PhD Student & Primarily Qualitative  \\
P17 & Sociology & Post Doc & Mixed-Method \\
P18 & Political Science & PhD Student & Mixed-Method  \\
P19 & Anthropology & PhD Student & Primarily Qualitative \\
P20 & Psychology & PhD Student & Mixed-Method \\

\bottomrule
\end{tabular}
\caption{Participant Summary Table. Participants self-reported their academic domains, position, and approach to research.}
\label{tab:participants}
\end{table*}
\end{footnotesize}

The interview protocol was designed to elicit perspectives on LLMs in qualitative research, with a focus on the data collection and analysis processes.
% , as also demonstrated by other work in qualitati~\cite{jiang2021supporting}.
% as they are more unique to qualitative research than other steps of the research process, which have been the focus of other work~\cite{kapania2024m}.
For this study, data collection was defined as \textit{any part of the research process that contributes towards how researchers obtain data from participants.}
% LLMs may play an active role in data collection, e.g. through creating recruitment materials for different formats.}
Data analysis was defined as \textit{any part of the research process that relates to the analysis of gathered data.}
% LLMs may play a role in data analysis, e.g. through applying qualitative codes.}
We opened the interview with questions about the researcher's qualitative work, practices, and their general familiarity with large language models.
We then asked questions about data collection and analysis, asking the researcher if they have or could imagine using LLMs for any portion of these research processes.
If they said they had used LLMs for that purpose, we asked them about their motivations for doing so, experiences of doing so, including perceptions of model performance, and ethical considerations during the process.
% how well the LLM performed, any ethical concerns they had, and whether they would use an LLM again for the same purposed.
If they had not used LLMs for the steps we outline, we asked them why they had not, and whether they had concerns with others using LLMs for these purposes.
% In both cases, we asked researchers to walk us through specific research studies where they did (or did not) use an LLM, to elicit precise discussions.
% Following this set of questioning, 
We concluded by discussing the participants' overall perceptions of LLMs in qualitative research, and soliciting both their perceived potential use cases as well as their concerns.
% , and concerns from incorporating LLMs in qualitative research.
Interviews lasted 42 minutes on average (min: 24, max: 59).
The full interview protocol is included in the Appendix.

% Once we had collected interview data, we set about 
We iteratively analyzed the data using inductive thematic analysis techniques inspired by grounded theory~\cite{braun2006using}.
While data collection was ongoing, three researchers open coded two interviews each, and each compiled a proposed codebook centered around identifying main themes, which were consolidated into one hierarchical codebook.
These initial themes motivated us to balance recruitment between computing and non-computing scholars, and participants with varying approaches to qualitative research.
% One researcher synthesized all three into a single codebook, which the rest of the team provided feedback and iterated on.
Data collection continued until we obtained a balance of academic domains and research approaches, and the main identified themes remained stable indicating theoretical saturation~\cite{glaser2017discovery}.
Using our first-round codebook as a starting point, two further interviews were fully coded by three researchers each, and the codes were refactored to yield the final codebook.
% Two further interviews were fully coded with this first-round codebook, and we iterated once more on the codebook to yield the final codebook.
We then divided the twenty interviews among coders, and met regularly to discuss the process and update codes where needed. 
The two primary authors re-reviewed each transcript and the applied codes for thoroughness, verifying with each other when uncertain.
Throughout the analysis process, we also constructed various artifacts, including interview notes, memos, and tables that summarized key context about each participant.
% The two primary authors reviewed every transcript in the study to achieve saturation and provide context on all participants.
% Like much qualitative HCI research, we do not calcualte IRR since ``codes are the process not the product''~\cite{mcdonald_reliability_2019}.
We used Dedoose for collaborative code application to interview transcripts.
We primarily avoided using LLMs for this research, with the exception of finding specific works of related literature, which we then verified externally.
To protect the identity of participants, any specific research studies or topics have been changed.

We reflect on how our own positionality influenced this study.
The authors are mixed-method researchers in HCI, NLP, and communication.
In our own work, we have felt interest in incorporating LLMs into our research processes, while being unsure how to ethically navigate adopting this technology.
This work is an effort to consider these opportunities and challenges head-on and to shine a light on the value of qualitative work in HCI.

\section{Findings}

In these findings, we outline how the researchers we interviewed adopted LLMs into qualitative research, identifying their main risks and concerns, then concluding by reflecting on the fundamental philosophical tensions raised.

\subsection{Current and Potential Uses of Qualitative Researchers Adopting LLMs}

% \subsubsection{Current Adoption of LLMs into the Qualitative Research Process}

\begin{figure*}
   \centering
   \includegraphics[width=\linewidth]{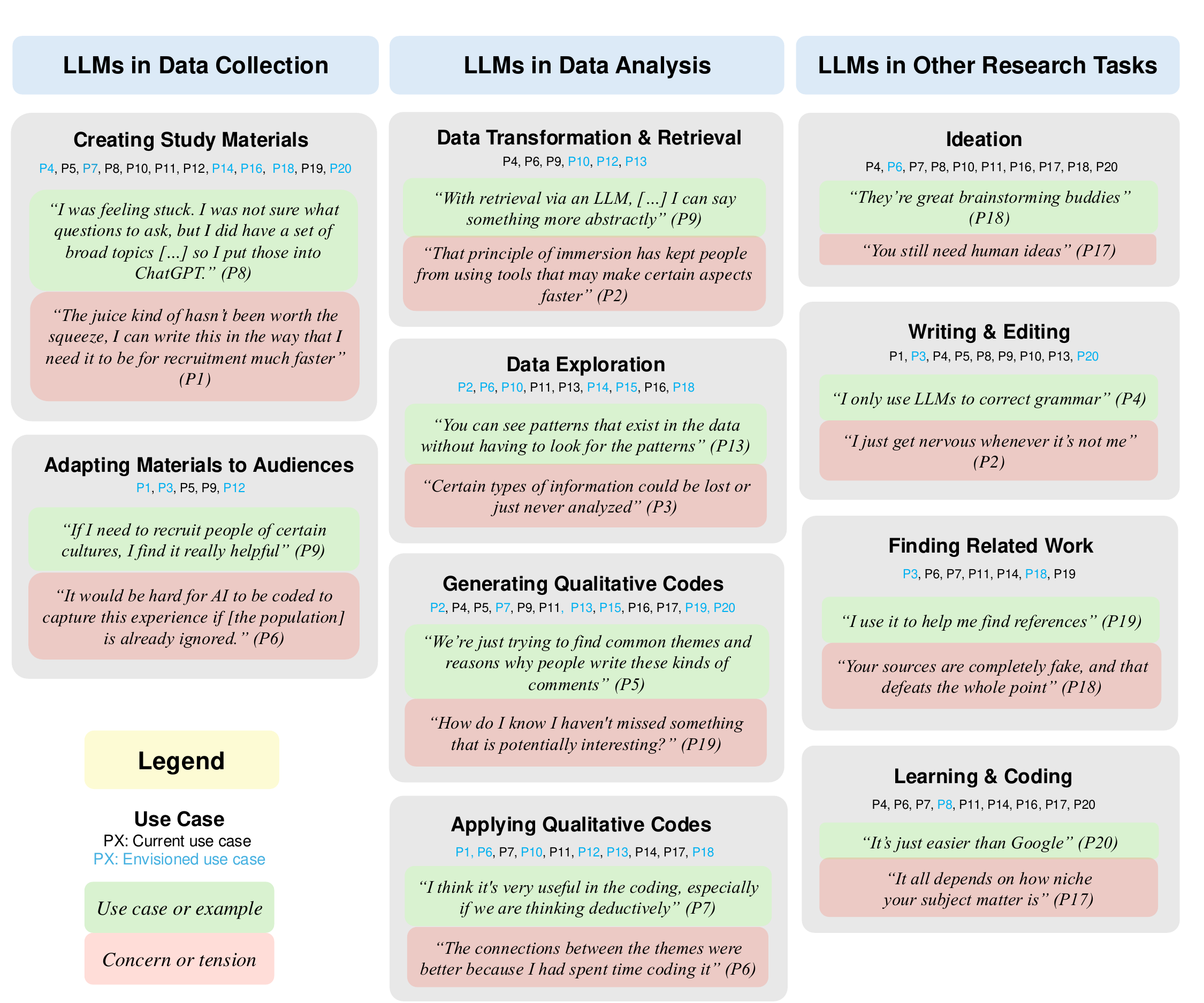}
   \Description{Current and Envisioned Use Cases for LLMs within the Qualitative Research Process. This figure highlights how participants had previously used and might imagine using LLMs moving forward in their research processes. To the left, the use cases under "LLMs in Data Collection" are Creating/editing study materials and Adapting materials to audiences. In the middle, the use cases under LLMs in Data Analysis are Data Transformation \& Retrieval, Data exploration, Generating qualitative codes \& themes, and Applying qualitative codes. In the right, the use cases under "LLMs in Other Research Tasks" are Ideating, Writing \& editing, Finding related work, and Learning \& Coding. Each use case is accompanied by a participant quote that supports and refutes the use case.}
   \caption{Current and Envisioned Use Cases for LLMs within the Qualitative Research Process. The figure highlights how participants had previously used and might imagine using LLMs moving forward in their research processes. We also highlight different perspectives and tensions for each use case, with participant quotes. Participants are only tagged under a use case if they view LLMs as being potentially beneficial to that end, and use cases are only included if mentioned by three or more participants.}
   \label{fig:qda_uses}
 \end{figure*}

% In this section, we report on both current use of AI tools
We sought to understand how our participants had or had not experimented with using LLMs in their qualitative research process.
Of the researchers we interviewed, fourteen reported actively exploring the use of LLMs in data collection or analysis. 
The most common use cases mentioned, highlighted in Figure~\ref{fig:qda_uses}, were using LLMs to help generate recruitment materials, attempts to use LLMs to speed up qualitative coding, and using LLMs for ideation and feedback. 

ChatGPT was the most frequently mentioned LLM tool by participants.
% Particularly for the scholars who w
When we opened the conversation by asking participants about their experiences with ``generative AI tools or large language models,'' most would immediately talk about ChatGPT.
Some researchers specified the model they used; P11 opened with \textit{``I use LLMs, specifically ChatGPT.''}
% In contrast, 
Others spoke more hesitantly, such as P13: \textit{``I've started using ChatGPT, I'm trying to be more familiar with it because it is the future, and you don't want to be a dinosaur.''}
Overall, perceptions of LLMs writ large were strongly influenced by experiences with ChatGPT as the most popular interface through which to explore LLMs.
Participants felt that LLMs were more advanced in areas where ChatGPT performed better, thanks to their interactions directly through the web interface. For example, more participants felt that LLMs were helpful \textit{``brainstorming buddies''} (P20), than felt they were helpful annotators, a use case for which ChatGPT's web interface was not designed. 
% Fewer of our participants had considered this or used LLMs as annotators.

% breaking up ChatGPT from qualitative tools and other types of tools
Outside of ChatGPT, participants shared that they used a suite of available tools for qualitative research, such as NVivo, Atlas.ti, MAXQDA or Dedoose, and AI sometimes underpins these tools' features.
% Participants frequently mentioned specific qualitative data analysis software such as NVivo, Atlas.TI, or Dedoose.
% Participants 
Participants were sometimes unsure whether these tools leveraged LLM-based functionality.
For example, participants (P4, P8, P9, P13) wondered: \textit{``does Grammarly count?''}
% For example, many participants asked whether Grammarly is considered an AI tool.
Others mentioned explicitly that they engaged in AI features offered by qualitative coding platforms. For example, P7 reported using the in-context code suggestions that Atlas.ti provides.
Participants did not necessarily feel agency over the AI features embedded in these tools, as P1 explained: \textit{``Atlas sends me a thematic analysis that essentially aggregates stuff, I don't really like it,''} and P4 summarized AI features in these tools as something they see \textit{``not by choice.''}
% P4 explained that \textit{``not by choice, sometimes the software will produce a list of bullet points.''}
% Similarly, P1 
Overall, participants often relied on software platforms to help with qualitative research, though not all appreciated the apparent increase in ``smart" AI-based features.

In contrast, some researchers shared they purposefully avoid advanced technology in their qualitative research process.
While all participants used their computers for analysis, they also described paring back their approach to remain close to the data, particularly to data from participant interviews.
P6 explained that while they were \textit{``not quite as old school as the people who would cut out their coded texts and move it around, I basically do that in Word documents.''}
%  [...] I don't even use a qualitative management software.
P2 echoed this sentiment: \textit{``I didn't like the restrictions and things the platform was putting on me, and I got really frustrated; I'm just familiar with Excel.''}
These attitudes highlight how, to some researchers, complex software interferes with their process or can be hard to learn, and simpler solutions may meet their needs.
% also P19 quote if needed: My typical process is that I go through the interviews myself and it's very old school I highlight in different color. Yeah, okay, You know, that comes up. Yeah. And then I reorganize what I like to put together. Like I regroup them. Yep. Have I ever used an LLM for that? No. Why? Well, I probably know why it's because I figure out the themes as I read.
% These sentiments seemed particularly prevalent in non-HCI fields.

\subsubsection{LLMs during Data Collection}
% A more specific common use for LLMs in the qualitative research process was to help researchers create recruitment and interview materials.
Seven participants 
% (P5, P8, P9, P10, P11, P12, P19) (in figure)
reported using LLMs to help with creating or editing recruitment or interview materials.
% their recruitment or interview materials.
Often, this assistance took the form of researchers using a service like ChatGPT to edit or tweak their writing.
Even those who were reluctant to use LLMs for research conceded that they could be helpful for writing, such as P10 who said \textit{``the only real reason that I use ChatGPT is to have it reword or rephrase,''} and P4 explained that \textit{``it will give me a really good sentence or switch a word.''}
This capability was particularly helpful for researchers whose native language is not English; for example, P8 said they \textit{``feel more confident once [they] have passed [their] writing to the AI tools.''}
LLMs also assisted participants in tailoring recruitment materials to their subject population. 
P9 (a non-American, non-native English speaker) 
% One notable example was P9, a non-American researcher whose first language is not English, who
told us they used LLMs to make messaging more appropriate for American research participants. They shared, \textit{``one thing I noticed that LLMs can do beautifully is make things American,''} and therefore \textit{``an LLM's general answer or general email to an American future participant will probably be better than my take on it.''}
% Similarly, P19 explained that they will actually test interview questions from different perspectives, for example asking LLM models to ``answer the question as if you're white, young, middle class, or answer the question as if you're Indian and working class.''
P5 reported that LLMs helped them deploy recruitment materials in multiple languages, explaining that they \textit{``used it for the translation of interviews and consent forms and other required documents,''} before having a native speaker collaborator double check the output.
% I did use large language models for the translation, and then I double checked it myself and with a native speaking collaborator double checked it as well, so
LLMs were also used by participants to help them create interview and recruitment materials from scratch.
For example, P8 recalled \textit{``I was working with a deadline, and I was feeling stuck. I was not sure what kinds of questions to ask, but I did have a set of broad topics [...], so I put those into ChatGPT and I was like, `Can you help me come up with this type of interview question?'''}
While quite a few of our participants had experimented with using these tools to help generate research materials, not all found these tools helpful: \textit{``the performance was really great,''} P5 said, while P8 thought \textit{``I feel like I  could have done better,''} and P10 elaborated that its outputs were \textit{``really broad, versus the empirical work I've done is very specific and granular.''}

In terms of promising potential uses, researchers seemed to understand why one might use LLMs to help create recruitment materials.
Some experienced researchers explained that they had past recruitment templates already, but for those who do not, P10 said, \textit{``I can imagine that maybe using something like this would be helpful for beginner researchers.''}
P20 expressed a similar attitude: \textit{``I would consider it in the future, if I'm drafting a new consent letter or things like that, but we had all of that from previous ones.''}
Nonetheless, some researchers pointed to specific reasons why they may not consider LLMs appropriate for these use cases, such as when materials are being produced for at-risk or non-Western populations, issues which will be further discussed in the following sections.

\subsubsection{LLMs during Data Analysis}
Another avenue that participants had explored, though often less successfully, was using an LLM to process and make sense of their qualitative data.
Many researchers we interviewed did not attempt to use LLMs in data analysis due to concerns about privacy and performance, 
% , since they struggled to understand whether they could share even anonymized participant data with third-party LLMs, 
issues we explore in the next section. 
Among those interested in LLM-facilitated analysis of qualitative data, we saw various functions being performed with LLMs: transforming or retrieving data, data exploration, generating initial codes or themes, and applying qualitative codes.

Researchers wanted to use LLMs as assistants for retrieving, processing, transforming, or reformatting their data using natural language. 
P9 reported that they leverage LLMs to perform fuzzy searches on qualitative data: \textit{``With retrieval via an LLM, I can say something like, `Where in this document is an example of such and such?' and just talk more conceptually or more abstractly, and the tool will find it.''}  
Participants also saw LLMs as being potentially useful to reformat data. For example, P4 imagined that \textit{``it would be nice if there was like a large language model agent to grab all the qualitative codes [and put them] onto a Google Jamboard because I have to rely on copy pasting.''}

On the analysis front, some participants explained that they valued being able to use LLM-aided software to flexibly explore participant data.
In this data exploration use case, participants reported using chat-based models like ChatGPT to conduct initial sensemaking with and about their data.
In P15's words, the ability to easily analyze qualitative data with an LLM was a \textit{``game changer,''} since \textit{``there's no friction between the person trying to make sense of a lot of data and the information.''}
% Overall
% Specifically, Y participants had previously attempted to perform a thematic analysis or assign qualitative codes to texts or transcripts.
One participant said they once used ChatGPT to help them make sense of the theory behind their qualitative data, explaining that they had interviewed people about their experiences with grief. They were \textit{``getting all this lovely, well-gelled data where people were repeatedly saying the same things, everything was perfectly categorized"} but the researcher was \textit{``having a hard time naming those categories in a way that was as inclusive as it could be.''} 
In this example, the participant asked ChatGPT to come up with a list of potential themes, based not on their participant data but on prior theories, and it gave them \textit{``this list of things, and I was like, hmm, this is really neatly packaged [...] it's giving me labels for the things I'm seeing that are also very intuitive.''}
These themes then served as a jumping off point for deeper engagement with the data, and guided the researchers' thinking for constructing their final qualitative codebook.

Experiences were more mixed for participants who directly tried to use LLMs to generate or assign qualitative codes to data.
In these use cases, participants were seeking to replace or augment the process of generating and/or applying a codebook to label qualitative data.
The most common experience was perhaps that of P5, who \textit{``tried to use ChatGPT to get the very general theme and the results were not that great, so I decided to code it manually, but I tried it.''}
% Some participants tried to use ChatGPT for generating codebooks or themes, with generally mediocre results(\textit{``vague''} (P6), \textit{``too generic''} (P5), \textit{``very general''} (P13), or \textit{``uninteresting''} (P9)). 
P9 engaged in a process they refer to as \textit{``semi-coding,''} where they provided ChatGPT a chunk of text they already had intuitions about, and wanted to see if the LLM agreed with their perspective.
Ultimately, however, they felt that the codes were \textit{``uninteresting in a way, or at least they were first level codes.''}
This experience was also shared by P11, who explained that when they \textit{``asked it to assign some initial codes [on their dataset], oftentimes the response will be very high level and abstract, so it's not trustworthy at all.''}
P17 disagreed, and had set up specific coding infrastructure that allowed them to identify whether an individual piece of social media data contains a rare occurrence of interest to their research.
P17 was seeking to identify and code specific types of gender-based online harassment in Reddit comments and derived the most advanced annotation pipeline we saw:
\begin{quote}
    \textit{``I come up with 16 or 18 thematic categories, and then use them to describe the system that I'm looking at. 
    That's qualitative work... I have... millions of comments, and most of those are garbage... So it's a needle in a haystack, and computation allows you to find and organize all those needles so that you can do the qualitative work.''}
\end{quote}
In their workflow, P17 derived codes based on initial data, and then constructed an LLM pipeline to annotate each post, using the LLM to take a first pass at organizing the data. 
P17 expressed that this process does not preclude deep qualitative work, since they still do a manual layer of qualitative analysis.

Despite mixed current experiences using ChatGPT for data analysis,  many participants felt curious or excited by the prospect of using LLMs to automate qualitative coding.
% (P6, P7, P10, P13, P18, P19). not necessary with the figure
The biggest factor that motivated the use of LLMs for annotation was the tedium and time investment required by manual coding.
% \textit{``I have thought about using it for coding,''} says 
P10 said \textit{``I'm currently in the process of coding, and sometimes I'm kind of like, it's taking a long time.''} 
Others mentioned that automating coding could limit human error and help them optimize for inter-rater reliability.
% and allow the coding to be more objective.
P13 coordinated a large project with multiple coders, and they described that \textit{``having a more thoughtful upfront conversation about that, and then using a tool to execute on that [code scheme], I think, would be way more accurate and better science than what we ended up doing.''}
Of those interested in using LLMs for coding, researchers had conflicting perspectives about whether LLMs would best be used for an initial, exploratory round of coding, or whether they should be used to apply manually developed codes.
% While not all researchers believed that LLMs could fully automate the qualitative coding process, 
Some specifically identified that LLM-aided coding could help with the first round of coding, or when initially getting a grip on the data.
P13 thought LLMs could \textit{``save you from that first tedious round of coding and then you could maybe be looking at the more nuanced codes faster over time.''} 
Others described the desire to use LLMs as a way to ``gut check'' their coding. For example, P18 explained that they had considered if \textit{``me and another human are coding the data, and then ChatGPT is also coding the data, and we do an intercoder reliability check with all of us.''}
% Overall, many participants saw potential for deploying LLMs to help them code qualitative data (P1, P6, P7, P10, P13, P18, P19).
% While many participants saw potential for deploying LLMs to help them code qualitative data, 
Nonetheless, some participants (P2, P4, P8, P16) struggled to see the automation of coding as aligned with their perception of qualitative work, stating \textit{``there's a part of qualitative research where it kind of feels like I'm in this for the manual work and that's why I'm doing it''} (P2). Some perceived LLM involvement in some research tasks as fundamentally in tension with qualitative research values, which we will further revisit in a later section.

\subsubsection{LLMs in Other Research Tasks}
Beyond data collection and analysis, participants also reported leveraging the interactivity of readily available LLMs, particularly ChatGPT.
Participants felt that ChatGPT could help them ideate better and be more creative researchers: \textit{``I think LLMs are really great for bouncing ideas off of and thinking creatively,''} P18 said, while P16 said they use LLM outputs \textit{``a basis or a starting point.''}
Participants also explained that they used ChatGPT to generate explanations for complicated concepts or coding implementations, 
like P7, who explained that they ask ChatGPT to help explain specific theories, stating \textit{``if I didn't understand what kind of methods they used in a particular paper or [...] I'm not able to make some connections or linkages, I just pose a question to ChatGPT.''}
In these instances, the social semblance of LLMs takes center stage, and our participants would sometimes refer to ChatGPT as more of a collaborator or peer than a tool.
For example, P18 referred to ChatGPT as a \textit{``great brainstorming buddy,''} P20 called it a \textit{``thought partner''}, P9 defined it as \textit{``a rational person, a smart layman, if you will,''} and P12 said that working with ChatGPT was \textit{``almost like working with another person.''}
For P11, this connection went further, as they found that interacting with ChatGPT while doing research helped them stay engaged with their research: 
\begin{quote}
    \textit{``As a researcher, we are always heads down and maybe we do not have much interaction with peer researchers who are human, I feel like I really value this kind of interaction, somebody I can talk to. I feel like this research process is not boring, it's engaging.''}
\end{quote}
    % I treat them sometimes as a friend who is knowledgeable across different fields and research areas, I feel like their ideas can be credible, and then we can have meaningful conversations. 
However, some researchers were more critical of ChatGPT occupying these kinds of interpersonal roles.
P5 expressed frustration that ChatGPT does not push back enough on their ideas to be a collaborator, since ChatGPT \textit{``never criticizes me, even though I asked: pretty, pretty please criticize me.''}
P4, meanwhile, was dismayed at the prospect of replacing any of their processes with ChatGPT, saying \textit{``I have a very supportive advisor and group of collaborators; they will give me great human feedback and I wouldn't I even think about using language models for that purpose.''}

\subsection{Perceptions and Risks of LLM Use in Qualitative Research}

While researchers identified potential use cases for LLMs in research, and expressed curiosity about new methods, most had not used LLMs extensively in research.
During our interviews, participants shed light on barriers to adoption, including a lack of established norms and concerns about output quality and appropriateness.

% so, what is driving that gap??
% can this be an equity question? between computational and non-computational fields?

% \subsubsection{Logistics and Startup Costs}
\subsubsection{Uneven adoption}
% \malq{not sure where to put this section?}
% ~\cite{messeri2024artificial}
Researchers reported logistical hurdles to adopting AI tools, which related to their comfort with technology and inequality around funding.
For example, participants (P1, P3, P10, P13, P18) expressed that adopting new technologies takes time and effort: \textit{``We're going to have to learn this, it's gonna take forever, and we're never going to get around to doing the actual project''} (P13).
These efforts could be even higher for researchers not in computing fields.
Researchers also may not be certain how to find tools that meet their needs; P20 described how \textit{``you don't know what's out there, sometimes you don't even know what something's called.''}
P2 and P5 recounted that cost influenced their choice of models, P2 stating \textit{``since the Claude model is quite expensive, I didn't use it, and OpenAI models are much  cheaper, so I could test it out.''} 
Overall, logistical issues led researchers to be more reliant on readily available interfaces like ChatGPT, even when these are not optimized for their use case, an issue exacerbated for non-technical researchers.

\subsubsection{Uncertainty about norms, guidelines, and expectations}

Participants shared that university or publication venue policies are shaping their perception of what is acceptable (P4, P11, P14), but many guidelines fall short. 
For example, participants referenced that university policies surrounding LLMs to produce writing informed their decisions about using LLMs for this use case, but also fuel uncertainty about using LLMs for other purposes. 
P14 reflected, \textit{``I guess, as students, ... it's kind of concerning that we might be penalized for using ChatGPT, so we're probably very, very careful''} and \textit{``I don't want this to backfire and cause any problems.''} 
% P2 elaborated: \textit{``I get nervous... whenever it's not me, when I feel like it's not me, is someone gonna say that I'm plagiarizing something?''} 
Sixteen participants in our study were graduate students, who occupy the unique role of collecting and analyzing study data directly, while feeling little power over the direct creation of guidelines.
Some researchers mentioned efforts to seek additional clarification, such as P13 who reached out to legal experts in the university to clarify whether they are allowed to \textit{``use AI''} when analyzing their data.

Some participants also raised concerns that LLMs introduce IP and originality issues that make them cautious about their use in research (P11, P14, P18, P20), particularly for tasks in research that create original  ideas or findings.
For example, P11 shared, \textit{``I feel like the IP is kind of mixed up... because ChatGPT does give me creative answers, but then sometimes I feel I just take it for granted, and I probably did not think about whether this creative answer has its own intellectual property.''} Opinions on acceptable use in light of this originality challenge were often inconclusive, with P19 asking, \textit{"Is it bad to ask it to do stuff, and then look it over, does that still count as your material? I think we're getting into those waters now."} 
Others note that as LLMs are increasingly integrated into workflows, policies that require disclosure, like the ACM policy, may be hard to enforce in practice. P18 wrestled with this concept, too: \textit{``At some point it may be somewhat impossible to truly distinguish what a human contributed or thought about, from what an AI that they were working with contributed.''}

\subsubsection{Navigating Participant Privacy Implications}
Our participants were most concerned with how LLM use impacts data security, consent, and obligations to participants in qualitative research.
Participants expressed a general lack of understanding and distrust for data protections awarded to participant data sent to proprietary models and AI companies.
For example, P7 ruminated: \textit{``how will the companies use that data [...] how is their data management protocol? Because the data that we share with LLMs, we don't know how that is managed at the back end.''} 

Researchers also wondered whether classic models of participant consent that cover sharing data with third party proprietary models like ChatGPT are doing enough to protect participants, like P8: \textit{``There's a bigger question of ... primacy of what is going to be giving consent.''}
% 
% like P9: \textit{``Am I allowed to upload even anonymized content to an LLM? Usually you don't sign consent specifically on that question. Maybe legally it's fine. But I feel like... it shouldn't be shared with a third party that might use it as training.''}
% While open-source model may be a potential solution to these questions, these were rarely brought up by our participants. (P18)
Some participants held strong positions on protecting participant data by not sharing it with proprietary LLMs.
P10 shared, \textit{``Obviously, I wouldn't use my participant data with GPT-4 and a lot of these other models where I don't know how they're taking care of the data.''} P18 expressed a similar strength of conviction: \textit{``I can't feed my data into ChatGPT, or something like that, because of confidentiality concerns.''}  
In the absence of clear, universal guidelines from IRBs within universities or venues providing recommendations for these issues, some researchers are creating their own strong norms: \textit{``I never put the data that is private to the model,''} P5 confirmed. 
% The researchers we interviewed were highly concerned about how LLM use might compromise participant data and privacy.

P7 and P13 highlighted that it is currently challenging to understand if LLM-based features are integrated into existing QDA tools in a way that protects participant data. For example, P13 noted that MAXQDA now has LLM-based features, but their team chose not to use them because they were unable to confirm the legality of doing so based on their data agreement and privacy standards. 
P7 echoed this, sharing their concern is not \textit{``just the researcher directly using ChatGPT, but also the researcher using tools which use ChatGPT further.''}
Researchers were also concerned these risks may be exacerbated for the populations of vulnerable groups qualitative research often seeks to study. 
Researchers in our study work with vulnerable populations, such as undocumented immigrants, those in political exile, and those with disabilities. 
The researchers studying these three populations shared explicit concern that using LLMs could put those participants' privacy at risk. Given the potentially grave consequences of identity exposure for individuals in those vulnerable groups, these researchers strongly felt those participants' data should not be used with LLMs readily available now.

To mitigate these risks, some researchers reported experimenting with ways to honor privacy while still using these tools. 
P11 shared, \textit{``Sometimes I do put some participant's quotations in, but I intentionally trim out specific keywords, for example, specific product names or functions.''} 
This participant is also an example of the multiple participants who reported believing it unethical to share participant data with third-party tools like ChatGPT, but who did not discuss potential alternatives, like open source models, that could address some of these concerns. This absence may be due to a combined lack of knowledge of privacy-preserving alternatives or the relative scarcity of mainstream, user-friendly alternatives to proprietary tools, particularly for users without technical backgrounds.

P9 said researchers should take a proactive stance towards participant privacy, particularly in the case of models that train on inputs: \textit{``Am I allowed to upload even anonymized content to an LLM? Usually you don't sign a consent specifically on that question. And maybe legally it's fine. But I feel like it shouldn't be shared with a third party that might use it as a training tool.''} 
Several proposed a stronger change to consent practices in anticipation of these concerns, like P7, who said \textit{``I think it's of the utmost importance to... let them know we would be following this method or doing this type of analysis wherein we are sharing some of their information with the LLMs.''}
% Broadly however, researchers appeared to follow their own norms, while feeling
Broadly, in the absence of strong guidelines, researchers grapple with appropriate use of these tools while still following their own convictions and research norms, leading to variable perceptions of acceptable and non-acceptable use.

\subsubsection{Concerns around validity and evaluation}
Participants considered whether LLMs would produce valid, usable outputs for research.
% \del{Participants cited that LLMs produce hallucinations and incorrect information. }
In particular, participants explained that the specific issues they feared or observed with LLM outputs included hallucinations, label assignment errors, or overly general responses.
\textit{``LLMs can give you wrong information''} said P18, and P4 agreed: \textit{``some points are just wrong.''}
More broadly, participants often found that LLM--- specifically ChatGPT--- outputs were vague, odd, or missing key points: 
% \textit{``it's not specific, it's vague, it's in some cases blatantly incorrect''} (P6); 
\textit{``they're getting some of these broad ideas, but maybe the variables or the concepts that are in this theory are a little off or they might be different theories,''} P6 shared, and P16 said
\textit{``the LLMs tend to write really generically and I would have concerns.''}
Researchers sometimes conclude these tools may not work for their qualitative research due to these limitations.
% , and many express concern that LLMs may produce something off, wrong or weird in the process of research.
% [cite] This underlies some reported distrust of tools we find in the participant group. [Insert quotes[]]
Researchers shared that the practice of collecting data (like interviews) themselves gave them an intimate understanding of the data, which could help them guard against misinterpretation of the data through an LLM's hallucination or misconstrual (P16, P20).

Researchers also reported uncertainty regarding the best way to validate and evaluate LLM performance in a way that aligns with their research community.
P18 shared that LLMs \textit{``are all constantly being updated [...] So the model is always changing, and that means that the output that you get is always changing.''} In response to these concerns about validity, many participants emphasized that checking the LLM's outputs is key to being confident in the use of these tools in research (P5, P7, P10, P14, P18, P19, P20) and some others also mention that it is unacceptable to not check outputs (P1, P5, P18).
P10, for example, emphasized the need for researchers to take responsibility: \textit{``there needs to be some sort of oversight because there will be an error or mistake through using LLMs, and it's on the researcher to catch those errors or to have some safeguard.''}

\subsubsection{Concerns about model bias}

Researchers raised concerns that biases embedded in LLMs might make them unsuitable for qualitative research on populations not well-represented in training data (P6, P7, P8, P10, P12, P16, P18, P19)
including non-English-speaking populations (P1, P12, P18, P19). 
Are LLMs \textit{``able to understand data from someone with a particular identity?''} P8 wondered.
Researchers reported both experiences with and fears of LLMs misinterpreting data from marginalized groups through the mistranscription of important language from that community, or by making judgments that are not faithful to the values in that community.  
For example, one participant noted \textit{``AI often falsely identifies queer content as hateful,''} confirming that an LLM's construals of their participant group may be at odds with faithful stewardship of that community's interests. 
Another participant noted ChatGPT gave reductive information about women from different backgrounds experiencing grief. \textit{``Probably, if you're saying, `what issues do women experience?' [ChatGPT is] going to try and give you the most broad, all-encompassing answer. There's not really any way to be sure that what it's giving you is covering all the ground.''}
P19 cited concerns over known Western bias as the main reason they do not feel comfortable using LLMs in their work:
\textit{``We all know that these models are pre-trained on datasets that are highly Western, mostly completely built from universities and companies that are based in developed places."} On their interest in using LLMs in qualitative research, they continued, \textit{``So for now, no, in the future I don't know, maybe. We have a long way to go for non-English speaking languages."} 
Researchers worried that these LLM biases could perpetuate existing power inequalities or harms towards marginalized groups.
P6 observed, \textit{``The machine is made within these power structures... so inherently it's going to privilege some things over others because that's how the world works,''} expressing a concern also shared by P12. Noticing similar properties of LLM training data, P7 said, \textit{``Most of the literature that's accessible is the voice of the powerful, right?''} 
Researchers were attuned to the connection between LLM training data and model outputs, and thus using LLMs may not reflect researchers' desire to responsibly analyze specific populations.
% Researchers were attuned to the connection between what data LLMs have access to and what their outputs can faithfully represent, a property creating a potential mismatch between the properties of LLMs and researchers' desire to faithfully analyze their subject populations.

\subsection{Qualitative Research and LLMs: Inherent Tensions?}
Simultaneous with curiosity and exploration of LLM tools described above, participants grappled with potential tensions between the values of qualitative research and the capabilities LLMs may enable. 

\subsubsection{Emergent insights from the data, versus interpretations imposed by LLMs} 
% Qualitative approaches stress uncovering insight from the bottom up whereas researchers are concerned LLMs may shape interpretation by presenting external, biased viewpoint.
Many qualitative researchers use methods where findings emerge from the data bottom-up, such as grounded theory, which encourages deep interaction with the data without imposing a view from pre-conceived theories. 
Researchers wondered whether bringing in the external interpretations, language, and potentially even theory from an LLM were in alignment with these qualitative approaches (P1, P3, P6, P9, P12, P16). P6 worried about letting LLMs lead theme generation: \textit{``My biggest worry is having AI code qualitative data: that feels a little bit different than engaging in a thematic analysis where we're letting the data drive the themes and we're recognizing that as humans, we're bringing the meaning to those themes.''} 
Using LLMs to create an interpretation of data could produce outright errors or produce standpoints that marginalize people. P16 said, \textit{"these people didn't agree to editing their words or modifying their perspective"} stressing the personal responsibility researchers feel in making sure participants are heard on their own terms. 
% On the other hand, past tools have supported researchers in exploring data in new ways \cite{jiang2021supporting} and builders could design tools to keep the values of exploration and emergence in mind.
A few participants reflected that relying on a single language model or sole output to engage with qualitative data might reduce the impact of their own perspective. P1 shared: \textit{``There's a little bit of like domain expertise that you're able to apply as a researcher and spend time in the space and spend time with these individuals,"} and in using LLMs to do this, P16 said, \textit{``I think we lose out on that, because now we're going with this very centralized, generalized model."} 

\subsubsection{Close engagement with the data, versus LLMs that automate processing} 

Participants reported that directly collecting interviews and reviewing them manually enables sensemaking and synthesis (P4, P6, P8, P10, P16, P19, P20).
P20 showcased this perspective, saying \textit{``I do think there's something in looking at all the transcripts, having done all the interviews yourself, and having intuitions around what themes arise.''} 
% (P20)
P10 also felt manual sensemaking was valuable, and tied this view to a potential tradeoff: \textit{``You also wonder: if an AI were to do it, would you lose some of that sensemaking that naturally happens when you're individually doing it yourself? And could that lead to not a strong findings or not a strong discussion section?''} 
Relatedly, some researchers worry that the ubiquity and ease of using LLMs in these tasks may lessen the degree of deep engagement with the data through manual sensemaking by allowing researchers to rely on "shortcuts." P19 shared that if someone is \textit{``using [an LLM] as a shortcut for reading large-scale things [where] you can't go through what's generated, then I do have concerns.''}
P13 saw the main risks of involving LLMs in data analysis as only resolvable if researchers still take care to collect and review data themselves: \textit{``I'm sure there's lots of ways this could get messed up, like only looking for certain words or words that go together and missing on nuance or interesting ways other people are talking about the same concept. If you'd never read it yourself... maybe something would get lost.''} 
Researchers wrestled with whether deep knowledge of the data would be maintained if LLMs were used in the analysis process.
% , echoing past concerns about technological involvement in qualitative research.

\subsubsection{Relationship and sense of responsibility to participants, versus an emotionless machine.}
Deep engagement with participant data, as well as an often ongoing relationship with research subjects and their stories, means that many qualitative researchers internalize a responsibility to do justice to participants in their research process and to act in their best interests. 
P18 expressed a fear about how LLMs could change this relationship: \textit{``If people found out that I was feeding data into an LLM, I think that can cause a breakdown of trust. Some of my participants are people I intend to continue working with in a more long term way.''}
When considering using LLMs for data analysis, P1 hesitated, saying \textit{``we're not necessarily building empathy in the same way, for participants''}, and P4 told us \textit{``it's not doing the data justice, you know, by just having an LLM look at it.''} 

\subsubsection{Subjectivity and multiplicity, versus one hegemonic way of seeing}
Our participants valued the unique context they bring to their own qualitative work (P1, P6, P7, P8, P9, P12, P16, P19, P20). Participants acknowledged interactions with LLMs may uncover new insights about their data, but they were concerned those interactions could result in some insights being lost.
For example, qualitative researchers see value in noticing things beyond the surface of transcribed interview text (P4, P6, P7, P10, P14, P16, P20), and note that LLMs do not see or know what humans do.
P9 shared, \textit{"I think you do gain a lot of insights by not just the words that are typed: the way I hesitate on certain things like accents, body movement... there's a tacit knowledge of the qualitative process that is not easily formalized and programmed."} 
As P7 put it, qualitative researchers \textit{``situate the text in a context,''} and do not just analyze the \textit{``text as it is.''} This led P7 to conclude that \textit{``in such cases, the LLMs might not be very helpful.''}

% As helpful as LLMs may be in discovering patterns or deriving new insights from the data, r
Researchers note that LLMs may change what humans notice in their data, and could also potentially miss critical insights. P19 noticed:  \textit{``There's just something interesting that I see that it doesn't give me, and the most important point of my analysis is to figure out the themes. If they're figured out for me, how do I know that I haven't missed something that is potentially interesting?''} 
% Echoing a quandary regarding what an LLM would surface in data, P14 said \textit{"I would look for a different weighing of priorities."}
% If researchers rely on LLMs to analyze data, and LLMs shape what humans notice in the data at the expense of insights humans would otherwise bring, some kinds of insight may be increasingly lost. 
P3 also wondered if LLMs would focus on different insights than humans, who collaboratively notice and integrate findings together: \textit{``I think when you ask an LLM to give you a specific set of instructions, you lose some of that nuance and some of that discovery. So I definitely think certain types of information could be lost or just never analyzed."}

\subsubsection{Who is in charge or creating meaning from human experience?}

Many researchers in our participant group discussed how qualitative methods are well-suited to understanding the \textit{why} behind human experience, (P1, P2, P6, P10, P13, P16, P20), particularly in contrast to what quantitative approaches can tell us. 
One theme is resounding: qualitative researchers want to keep humans in charge and center their perspectives.
Summarizing this point, P8 shared, \textit{``The whole point of doing qualitative research is to engage with people, and if we just stop doing that, then I feel like it endangers qualitative research in some way.''} 
P15 took the view using LLMs in qualitative research may still be compatible with this goal as long as humans are the ones ultimately determining meaning from an analysis: \textit{"I think LLMs would be incredible for topical emergence. Only humans can really [tell] you what the themes are, ultimately."} 
% Can LLMs surface useful complementary insight while we keep humans in charge of what those features mean in the context of human lives? LLMs do not have lived experience, and do not ``understand'' the lived experiences of humans. In a set of methodologies concerned with making sense of the \textit{``texture of human lives''} (P19), keeping the project of creating meaning as a human-centered project is critical. 

\section{Discussion}
% Our interview themes allows 
Our interviews situate at least a partial answer on how qualitative researchers can leverage LLMs while staying faithful to responsible, participant-centered research.
In response to RQ1, we find that many qualitative researchers
are open to the conscientious use of LLMs across the research pipeline.
RQ2 poses the question of which tensions might arise if they are used, and we find that qualitative researchers harbor many
% In participant answers, we found optimism about many potential uses but also 
valid questions and concerns about how to align LLM uses to their personal and disciplinary research values.
RQ3 raises the idea of potential models for ethical use of LLMs in qualitative research, which we discuss in Section 6.

\subsection{From Task-Specific to Ubiquitous AI}
We build on prior work on human-AI collaboration in qualitative contexts~\cite{jiang2021supporting, tools_in_place} by situating AI use in the current moment.
\citet{tools_in_place} argued that successful human-AI collaboration for qualitative research requires the intentional use of AI for specific tasks, such that researchers are not ``displaced as the primary analysts.''
~\citet{jiang2021supporting} similarly suggest future research explore which precise sub-tasks of qualitative research are most appropriate for AI involvement.
Additionally, many papers on human-AI collaborations reported that the order of AI involvement matters in qualitative research, since researchers do not want to be influenced by suggestions too early in their bottom-up sensemaking process ~\cite{lam2024concept, overney2024sensemate, jiang2021supporting, tools_in_place}.
Aligning with this idea, our participants were comfortable with task-specific tools that use LLMs such as grammar check, speech-to-text transcription, and translation, and used them extensively.

However, our findings highlight that the ideal of intentionally using specific AI tools for clearly delineated research tasks is increasingly challenging to meet, due to the prevalence of AI through flexible, seemingly all-purpose chatbot interaction patterns popularized by ChatGPT.
Participants used AI across their research pipeline, a finding also supported by ~\citet{liao2024llms}. 
A natural consequence of this shift is that the lines between AI and non-AI tools blur, making intentional  use more challenging.
~\citet{narayanan2024ai} speculate that generative AI will become ``part of our digital infrastructure, instead of being a tool people use for specific purposes.''
Our findings suggest that this shift is underway.

Nonetheless, this shift need not be solely negative.
Our participants indicated that the ``helpful assistant'' nature of the most popular AI tools today collapse the boundary between seeing LLMs as a ``tool'' and as a ``collaborator.''
% Nonetheless, there are also positives of this paradigm: for example, the interactive, 
While there are significant risks of overly anthropomorphizing AI~\cite{messeri2024artificial}, these natural language interactions may also be the kind of human-AI collaboration that ``support serendipity,'' which ~\citet{jiang2021supporting} champion.

\subsection{The Sociotechnical Breakdown of Guidelines, Norms, and Tooling}

Worries about adhering to norms and guidelines are ubiquitous, despite the active development and publication of norms by institutions and publication venues (e.g. ACM policies), ethical and supervisory boards, and academic research papers~\cite{watkins2023guidance, hosseini2023fighting}.
In the social norm change literature, scholars posit that there is often a tipping point where social norms change rapidly~\cite{andrighetto2022research}.
Our research suggests that some norms regarding LLMs may be close to a clear tipping point, such as the understanding that identifiable participant data ought not be sent to a third-party tool that may train AI models on the data, while norms regarding some other concerns have so far eluded convergence.
Our work confirms prior perspectives that AI evolution is outpacing our norm formation processes~\cite{baronchelli2024shaping, feldstein2024evaluating}, and we show that scholars feel pressure and uncertainty when justifying LLM-aided techniques for academic publications.
% Other work suggests that there may be systematic divides between fields in who is most comfortable disclosing use of LLMs~\cite{liao2024llms}.

We show that LLMs are being broadly adopted in qualitative research practices despite wide-ranging concerns about privacy.
Ongoing work suggests that non-computing researchers have more pronounced ethical concerns in using LLMs~\cite{liao2024llms}, and our findings suggest that ethics and privacy were top of mind for qualitative researchers more broadly.
Nonetheless, the present research suggests that participants use or consider using LLMs \textit{despite} these concerns, and sometimes use these tools while still being unsure about appropriate use.
~\citet{kapania2024m} similarly find that researchers are often aware of potential ethical issues, but uncertain about how to address them.
% We observe among our participants that these fears are likely to be compounded for junior scholars.
We position the current research environment as a fundamental sociotechnical breakdown: the available \textit{tools} do not present researchers with the right options to confidently preserve privacy, the necessary \textit{guidelines and norms} are lagging behind, and the \textit{social context} is increasingly normalizing and even expecting the use of AI tools.
This misalignment of incentives resembles Ackerman's notion of the socio-technical gap~\cite{ackerman2000intellectual}, which refers to the gap between technical capabilities and social expectations.

While the imagination of the majority of researchers we spoke with was shaped by the current capabilities of ChatGPT, the most technologically savvy participants were currently leveraging LLMs to do more complex tasks including systematic qualitative coding.
Several participants wanted to use LLMs to do qualitative coding, but did not know how to use current tools, indicating their lack of use now could be partly due to a skills gap.
% rather than purely ethical concern.
% This digital divide indicates that participants' technical familiarity currently limits what they deem possible.
Thus, the deployment of intentionally designed tools like CollabCoder \cite{gao_collabcoder_2023}, SenseMate \cite{overney2024sensemate}, and TIQA \cite{ tseng2025mitigating} is even more important, since the state of accessible tooling may be actively limiting responsible research.
We also see evidence that this divide in adoption could, on the other hand, be more fundamental: some participants simply deemed AI tools for tasks like coding to be needless departures from their qualitative sensemaking process.

\subsection{Fundamental Tensions in Using LLMs for Qualitative Work}

% \del{Beyond these practical concerns,} 
A set of potentially fundamental tensions arose in considering the use of LLMs to conduct qualitative research.
There has long been a tension between work seeking to combine computational power with interpretivist approaches \cite{baumer_comparing_2017,nelson2020computational} and work arguing that qualitative values must be protected from quantitative evaluation standards~\cite{crabtree2024h, soden2024evaluating}.
Our interviews suggest that LLMs may continue blurring the line between positivist and interpretivist approaches, as exemplified in emerging work~\cite{lam2024concept, choksi2024under}. 
As~\citet{messeri2024artificial} discuss, the way some approaches are more readily aided by LLMs may change the research landscape by making some questions more tractable than others.
% We also see glimpses of this cleave in the digital divide between 
There is a danger that the ability to analyze data at scale with less human involvement may make situated, slow, contextual and interpretivist approaches to qualitative data less attractive, potentially because of incentives to quickly publish, or because of the perceived objectivity of LLMs compared to human analysts.
Mixed-method researchers, who make up roughly half of our participants, may be particularly positioned to accelerate this change. Published work uses LLMs for many types of text analysis, including in powerful new tools that are increasingly positioned to facilitate ``theory-driven analysis" of textual data~\cite{lam2024concept}, and these tools may be adopted by users from different epistemological traditions.
As LLMs produce summaries, topics, thematic interpretations, and labels of textual data that look increasingly fluent, nuanced, and contextual, researchers may increasingly be tempted to use these outputs in analysis while decreasing the type of engagement with participants and their data that researchers say is central to qualitative research values.

\section{Implications for Researchers and Designers}

\subsection{Considerations for researchers using LLMs in qualitative work}

Thus far, we have explored how the combination of opaque tooling, lagging norms, and social context lead to an uncertain environment for qualitative researchers today.
In this section, we answer RQ3 by providing an overview of key considerations and recommendations for LLM-curious qualitative researchers, or those who feel that LLM tools may interfere with their processes.
Maintaining researcher agency and educating researchers on AI tools is vital as LLMs are embedded across more surfaces.
% As AI tools become harder to use intentionally, maintaining researcher agency in tool use becomes more important.
% We use prior work and our own expertise to reflect on the tensions raised by participants.
We used prior work and our own expertise to reflect on the tensions raised by participants; we thus constructed Table~\ref{tab:choices} to distill some main choices researchers can make relating to the uses and potential implications of LLMs in qualitative contexts.

\begin{footnotesize}
\onecolumn
\begin{longtable}
{|p{0.15\textwidth}|p{0.375\textwidth}|p{0.375\textwidth}|}
\caption{Table highlights the main considerations and recommendations regarding how qualitative researchers can incorporate LLMs into their research process. \textit{Considerations} include a brief overview of the main tradeoffs involved in specific decisions, while \textit{Recommendations} provide explicit guidelines for researchers that minimize risk to privacy, sensemaking, and research ethics.}
% \label{tab:choices}\\
%     \hline
%     \textbf{Category} & \textbf{Considerations} & \textbf{Recommendations} \\ \hline
%     \endfirsthead

%     \hline
%     \textbf{Category} & \textbf{Considerations} & \textbf{Recommendations} \\ \hline
%     \endhead

    % \renewcommand{\arraystretch}{1.5} % Increases row height (vertical padding)
    \label{tab:choices}\\
    \hline
    \textbf{\Large \rule{0pt}{10pt}{Category}\rule[-5pt]{0pt}{8pt} } & 
    \textbf{\Large \rule{0pt}{10pt}{Considerations \rule[-5pt]{0pt}{8pt}} } & 
    \textbf{\Large {\rule{0pt}{10pt} Recommendations}\rule[-5pt]{0pt}{8pt} } \\ \hline
    \endfirsthead
% \label{tab:choices}\\
% \hline
% \textbf{\Large \hspace{5pt} Category \hspace{5pt}} & 
% \textbf{\Large \hspace{5pt} Considerations \hspace{5pt}} & 
% \textbf{\Large \hspace{5pt} Recommendations \hspace{5pt}} \\ \hline
% \endfirsthead

% \renewcommand{\arraystretch}{1} % Resets to default row height for the rest of the table

    \hline
    \textbf{\Large Category } & 
    \textbf{\Large Considerations} & 
    \textbf{\Large  Recommendations} \\ \hline
    \endhead

    \hline
    \endfoot

    \hline
    \endlastfoot
    \multicolumn{3}{|l|}{\textbf{\normalsize \rule{0pt}{10pt}Policies \rule[-5pt]{0pt}{8pt}}} \\ \hline
   \rule{0pt}{10pt}Institutional, venue, and government guidelines and regulations \rule[-5pt]{0pt}{8pt}& Policies and regulations can help to provide basic guidance, but may be out of date or insufficient by themselves. &
    Familiarize yourself with and follow institutional, publication, and federal policies, but treat these as a starting point for ethical engagement. \\ \hline
    \multicolumn{3}{|l|}{ \normalsize \rule{0pt}{10pt} \textbf{Research Ethics \& Values} \rule[-5pt]{0pt}{8pt}} \\ \hline
    \rule{0pt}{10pt} Consent \rule[-5pt]{0pt}{8pt} & Participants may not know that their data could be shared with a third-party company, or be aware of potential implications, such as the implication of models training on their data. &
    Update consent forms on details if planning to use LLMs in the research, particularly in a training capacity, and educate participants on potential risks. \\ \hline
    \rule{0pt}{10pt} Privacy \rule[-5pt]{0pt}{8pt} & Guidelines may not strictly disallow using participant data with proprietary LLMs, even ones that train on inputs, leading to identification risk in the future. Additionally, LLMs may be embedded in the tools you currently use. &
    Do not use proprietary models that train on input when interacting with participant data. Host local models and use open source models when possible. Anonymize confidential participant data if using a proprietary model. Confirm data policies before using a tool like MAXQDA that integrates LLM features. \\ \hline
    \rule{0pt}{10pt}Disclosure \rule[-5pt]{0pt}{8pt}& Venues may mandate specific forms of disclosure, but researchers may wish to disclose additional uses of LLMs for research. Reflexively engage with how LLMs may have shaped the relationship with participants, ideas in the project, and perceptions of the data. &
    Proactively disclose LLM uses that significantly impacted a sensemaking process or anaysis outputs, including in ideation. When LLMs are used in analysis, discuss validation, including, disclosing model, model version, prompts, and performance, if appropriate.  \\ \hline
    \rule{0pt}{10pt}Intentional task selection \rule[-5pt]{0pt}{8pt}& Chat-based interactions make unintentionally transitioning from one task to another easy. Tasks carry heterogeneous benefits and risks for qualitative research, and some decisions about LLM involvement or automation depend on research values. Interpretivist qualitative research tasks may be particularly at odds with LLM use.  &
    Before using an LLM, outline all tasks under consideration for LLM use. For each one, decide if the task is appropriate for LLM involvement or automation based on risks to compliance, originality, bias, privacy, and validity according to research values. \\ \hline
    \rule{0pt}{10pt}Validation \rule[-5pt]{0pt}{8pt}& LLMs should always be validated in task-specific ways in research.
    % If LLMs will be part of your research process, they should be validated for that task. 
    A clear way to validate the performance of an LLM may or may not exist for every task, and some tasks may be inappropriate for LLMs to perform even if they can be validated. Formal validation may be more tractable for positivist research, and a plan for validation should be made prior to engagement. Interpretivist and other research stances should consider whether any tasks are appropriate, given challenges to validation. &
   Decide how to validate an LLM’s performance on any chosen task. This could include side by side comparison to human performance in qualitative coding, for example, or manual review of any edits to text suggested by an LLM. Only use an LLM if performance can be appropriately validated for the task. \\ \hline
    \rule{0pt}{10pt}LLM bias\rule[-5pt]{0pt}{8pt} & Consider LLMs' cultural and political bias as you decide which tasks are appropriate for an LLM. Consider whether  participant populations are aligned with LLM training data, and what the impacts of misalignments might be.
    &
    For studies on populations unlikely to be well-represented in training data of off-the-shelf LLMs, avoid using off-the-shelf models in analysis to limit misinterpretation, misrepresentation, or hallucination, or consider fine-tuning. \\ \hline
    \multicolumn{3}{|l|}{\normalsize \rule{0pt}{10pt} \textbf{Tool Selection} \rule[-5pt]{0pt}{8pt}} \\ \hline
    \rule{0pt}{10pt}LLM chat interfaces \rule[-5pt]{0pt}{8pt}& Chat-based interactions are highly interactive and can support brainstorming in a research process, but interactivity can induce slippage across tasks and be difficult to validate for every use case. As such, some researchers may decide the risks of task-agnostic chat interfaces are not worthwhile. & If using a chat interface, clearly outline the tasks for which the chat-based LLM will be used, and make a plan for validation before approaching the task. Monitor regularly for slippage across tasks. \\ \hline
    \rule{0pt}{10pt}QDA Software \rule[-5pt]{0pt}{8pt}& Software for QDA is intentionally designed for a qualitative coding use case, so it may provide validation (like measuring IRR across collaborators) or be designed with risks to sensemaking in mind. However, highly structured tools may miss opportunities for interactivity afforded by LLMs, so appropriateness depends on the intended task. & Using structured tools reduce some but not all risks. If using LLM integrations into existing or emerging software tools, investigate the platform's privacy policies and still consider the bias an LLM could introduce into your analysis. \\ \hline
    \rule{0pt}{10pt}Task-specific tools (e.g. Microsoft Excel, Grammarly)\rule[-5pt]{0pt}{8pt} & LLMs are increasingly embedded across the tools we use every day, even for simple tasks. Tools like Grammarly with more prescribed functionality constrained to editing may limit interference with your sensemaking process compared to open-ended tools. & Reflect on your full set of research tools. Monitor for potential compromise of participant data if you use Grammarly or other tools to process or edit participant data. \\ \hline
    \rule{0pt}{10pt}Non chat-based LLM use\rule[-5pt]{0pt}{8pt} & Researchers may wish to use or create a research infrastructure using an LLM, either directly or accessed through an API, for tasks like procedurally annotating data. In these cases, choices between proprietary models and open source models, or locally hosted and remotely accessed models become salient, and carry different implications for privacy, transparency, and quality. & If possible, use an open source model rather than a proprietary model to promote transparency and open science. If possible, use a locally hosted model for greater control over data. If using a proprietary model, anonymize data and do not use models or API services that train on input.
\end{longtable}
\end{footnotesize}
\twocolumn
We now explain the three main covered areas of Table~\ref{tab:choices}: \textit{Policies}, \textit{Research Values \& Ethics}, and \textit{Tool Selection}.

\subsubsection{Policies}

The first dimension in the table is the role of \textit{policies}.
We recommend actively keeping up to date on all policies from relevant communities, but treating these policies as an entry point rather than as a final answer on appropriate engagement with these tools.
Policies may vary between venues; for example, some venues mandate specific types of disclosure~\cite{noauthor_frequently_nodate} while others disallow the use of AI in some contexts~\cite{sagepolicy}. 

As surfaced by our participants, existing policies and regulations may not engage deeply enough with the risks to certain participant populations.
Therefore, while policies are important, they are likely insufficient considerations by themselves.
The rest of the table provides a scaffold for considering the conscientious use of LLMs in qualitative work beyond established policies. 
% compliance with IRB and venue policies that may not deeply engage with risks to participant data or originality of research may not be enough, making these important but insufficient considerations when engaging LLMs in qualitative work. 
% Researchers must balance 
% Participants
% When choosing to engage in disclosure, 
% If policies regarding disclosure of LLM use are followed, researchers may still be perceived differently, particularly in fields that may be skeptical towards LLM adoption in research. Researchers must navigate perceptions of the technology within their community above and beyond strict policy on usage. 

\subsubsection{Research ethics and values}

The second main dimension of the table encompasses \textit{Consent}, \textit{Privacy}, \textit{Disclosure}, \textit{Intentional task selection}, and \textit{Validation.}
Consent and privacy protection are fundamental values in qualitative research. Our participants indicated a current lack of scaffolding in current tools guiding how to use LLMs while preserving these values.
% , but existing policies may allow LLM usage with participant data while participants are uncomfortable with their data being shared with AI companies or used for model training. 
% Researchers should update consent forms so participants know how their data might be processed by LLMS, especially if used for training, as recommended by~\citet{kapania2024m}.
Researchers should update their \textit{consent} forms to accurately reflect intended uses of LLMs in research, as recommended by~\citet{kapania2024m}. 
These forms should include details on how participant privacy will be protected and education about potential risks.
% (through using anonymized data on local models, for example, in contrast to companies that may use participant data to train future models).
% Researchers may lean on the long hi

Participants repeatedly highlighted a fear of compromising participant \textit{privacy}.
Researchers must take care not to share personally identifiable information or compromise the data of at-risk populations.
These risks are particularly salient when using proprietary LLMs for any task involving confidential participant data.
% , so any decision to use an LLM must first take privacy preservation measures into account.
% , which is a risk of using especially proprietary LLMs in conjunction with confidential participant data. 
If using a proprietary model for such a task, carefully consider how to anonymize data, such as removing names or identifiable terms (e.g.~\cite{choksi2024under}).
Avoid LLM interfaces which use your participants' data as training data for the model.
Check whether any tools with LLM-based features (e.g. Grammarly or Atlas.ti) train on inputs, and how they handle data privacy before using them. Privacy may be better protected by hosting open source models locally, but this is not currently feasible for all researchers.
% that incorporates LLM-based functionality, check whether the tool is using a proprietary model that may change your assessment of its appropriate use with your data, either according to university policy or IRB standards. 
% As a matter of process, update your consent forms for participants to accurately reflect any intended use of LLMs in the research and implications for participants. 
% This should include details on how participant privacy will be protected (through using anonymized data on local models, for example, in contrast to companies that may use participant data to train future models).

Choosing which types of \textit{disclosure} are appropriate regarding AI use is an opportunity for qualitative reflexivity.
% Venues also have specific requirements regarding the appropriate use of LLMs in research and disclosure of their use, which should be followed. However, 
Qualitative researchers, as reflected in our interviews,
% occupy a unique position as LLMs change the social sciences: they are more likely than other groups 
% to already explicitly engage 
already center reflexivity and positionality in their work, and
% as a factor when considering the influences on their work. 
% Extending this idea, 
% Qualitative researchers 
% engaging in qualitative methods 
can set precedent by explicitly engaging with how an LLM shaped a project. 
Researchers could consider the potential benefits of additional AI disclosure beyond those required by a venue, institution, or community.
% Stigma 
However, disclosing the use of AI may risk stigmatization, especially in non-computing fields~\cite{liao2024llms}, which should be balanced with the benefit additional transparency and context disclosure about AI usage could bring.
% If policies regarding disclosure of LLM use are followed, researchers may still be perceived differently, particularly in fields that may be skeptical towards LLM adoption in research. Researchers must navigate perceptions of the technology within their community above and beyond strict policy on usage. 
% If you decide to use an LLM in these tasks, consider engaging in disclosure even if your community does not require reporting for your use case.
Disclosures should ideally detail which models were used, and any fine-tuning, optimization, or prompt engineering performed.

Our table also outlines the value of \textit{intentional task selection} when using LLMs in qualitative research.
Intentional deployment of AI is key to maintaining researcher agency~\cite{jiang2021supporting, tools_in_place}, but our participants highlight how general purpose tools now blur boundaries between tasks.
With this challenge in mind, intentional task selection is even more pressing.
% LLMs are being used for many tasks in qualitative research, but not all tasks have the same risks and benefits to researchers and participants. 
The appropriateness of involving an LLM in a task can depend not only on venue policies, but also on research values, like an interpretivist or positivist stance. 
Before using an LLM for any task in your research, researchers should outline all tasks under consideration for LLM involvement and determine whether involving an LLM is appropriate based on their research values. 
In particular, researchers should consider how involving an LLM in the task may shape their perspective, ideas, and findings if used in ideation, analysis, or writing. 
% If an LLM 
% If the role the LLM plays is central to deriving or informing core theoretical finding in their work, including ideation or informing analysis, the 
Researchers should also consider whether they may be committing accidental acts of academic plagiarism by relying on LLM-generated outputs~\cite{bhardwaz2023extensive} across the research pipeline, and should take responsibility for checking for copied content. 
If any tasks are determined to be appropriate for LLM involvement, researchers should monitor whether the planned task matches ongoing use of the LLM.  
% At the same time, there are increasing number of use cases where researchers may not have a choice in their use of AI tools: for example, as LLMs are increasingly embedded in surfaces like Google Search. 

% The \textit{validation} required to use an LLM 
Our table also centers \textit{validation} as a key consideration when using LLMs for qualitative research.
Whether and how LLM usage can be validated appropriately depends on the task under consideration and individual research values. 
For example, if using an LLM to edit text, an appropriate validation might be to personally review each LLM-generated edit. 
If using an LLM to scale data annotation beyond possibility of human review, it may become appropriate or even necessary to use more positivist measures of validation, like inter-rater reliability, to measure accuracy compared to a human baseline.
% appropriate validation of LLM annotation might be a direct comparison of LLMs to human performance on that task, a validation that might be satisfactory to positivist researchers. 
Not all qualitative research tasks have a clear appropriate validation: in particular, interpretivist researchers may find LLMs cannot be validated for tasks of interest and should be avoided. We advise making plan for validation of LLM performance for any chosen task \textit{before} starting to interact with the tool, since anchoring bias can cause one to rationalize findings and outputs after already interacting with the LLM.

Additionally, when considering LLM appropriateness for any task, we recommend considering \textit{LLM training data, current performance, and biases}. For example, LLMs may be useful for editing English text written by non-native English speakers, which appears to be a primary emerging use case from our participants and in other research~\cite{liao2024llms}.
However, LLMs may not yet be advanced enough in other languages to be used for the same purpose.
% research written in a different language may not be able to use current LLMs for the same purpose. 
% Participant interests should also be considered when weighing the 
Model performance on tasks like transcription and analysis often directly reflects training data~\cite{koenecke2020racial}, so alignment between participant interests and a model's training data should be key considerations.
% Also, participant interests and research values should be considered when weighing LLM appropriateness for a task. 
Using a mainstream LLM to analyze themes in a study of non-Western, non-English speaking participants may be inappropriate, since mainstream LLMs likely have poorer representations of those groups and viewpoints and could cause misinterpretation or harm.
% , in addition to lacking researcher context.
% We have seen similar things play 
% We have seen the harmful impacts of certain populations being underrepresented in training data~\cite{koenecke2020racial}.
If LLM capabilities and bias cannot be considered and confidently addressed for a potential task, the tool should not be used. 
% Consideration for LLM bias can be part of a researcher's validation of LLM appropriateness for a planned task.

% \subsection{Research process}

% Researchers are using LLMs across all stages of the research process, but involvement in tasks at different stages of the research process carries different considerations and risks. LLMs can be useful in ideation, but can can inspire derivative or wrong ideas about the data. For qualitative research methods like grounded theory which prioritize direct researcher saturation with the data to create meaning, LLM interference in a process of brainstorming insight from the data may be inappropriate. 

% If considering using an LLM for an ideation-related task, decide if doing so is aligned with your research methods, and if the way an LLM could influence ideas is in alignment with your research process. Engage in disclosure, as brainstorming with an LLM may shape original thinking on the data or how to organize main paper contributions.

% % LLMs can be useful aids in generating and editing study materials

% AI has already been used to support qualitative coding and collaborative data analysis. Using an LLM in analysis can support positivist annotation paradigms in particular. 

\subsubsection{Tool Selection} 

In the table, we outline the four types of LLM-aided tools that came up in interviews: \textit{LLM chat interfaces}, \textit{QDA software}, \textit{task-specific tools}, and \textit{non-chat based LLM use}.

Participants largely reported experiences with ChatGPT, which has interactivity and flexibility researchers enjoy but does not scaffold appropriate use for research purposes. 
Conversely, structured software solutions like MAXQDA for data analysis or Grammarly for text editing have LLM-based features but are designed to support users in a specific task-based use. 
Researchers should consider if a dedicated tool exists for their use case that guards against some risks to the user that open-ended chat interfaces pose. 
% If an LLM rather than a software interface with LLM-based features will be used, the interaction pattern the researcher chooses with the LLM is also a consideration. 
Chat is a useful way to interact with LLMs, but researchers may instead wish to integrate LLMs into software pipelines to procedurally generate outputs like data annotations.
% LLMs can also be integrated into software pipelines to procedurally generate outputs like data annotations.

The way an LLM is hosted and accessed also has implications for privacy and transparency. 
Proprietary models from companies like OpenAI can be accessed through a web browser or API, meaning data shared through those access points is shared with the company. 
Models from AI companies are often also closed source, meaning the model's training data and weights are private, and understanding model behavior can be difficult.
% making understanding training data and their behavior mechanisms difficult. 
If using a proprietary model for research purposes, consider providing explicit justification as recommended by ~\citet{palmer_using_2024}. 
Open source models with known model weights and transparent training data are preferable for transparency, data privacy, and control, but currently, hosting them locally can be difficult and time-consuming without resources and technical skills.

\subsection{Design principles for qualitative research tools incorporating LLMs}
% excellent quote from P19 if we can find a way to work it in: But I think the main issue is that people who build LLMs really need think about social questions the way we do it has to be from the beginning, not like after they're build and we try to like find loopholes, but I think they really have to be built with these socially conscious questions in mind, for us to be able to use them as qualitative researchers, use them in good faith.

We have identified the urgent need for transparent tools that center qualitative research use cases.
% In particular, our participants often relied on general purpose tools like ChatGPT for qualitative research tasks.
% While proprietary closed source models may not be approp
% showing the need for interactive paradigms 
% used chat-based proprietary tools like ChatGPT for qualitative research tasks, 
% Participants already use chat-based proprietary LLM interfaces like ChatGPT for qualitative research tasks, while dedicated QDA tools, do not currently serve comparable roles or have comparable features. 
% Conversely, general-purpose chat-based LLM interfaces are not designed as qualitative research tools.
% We outline key opportunity areas for the development of tools for qualitative researchers that leverage LLMs.
% We note that tools built with these findings may not be beneficial or appropriate for all qualitative researchers, particularly those who choose to eschew the use of LLMs in their work for any reason, including conflict with their epistemic stance or risks related to bias.
% Based on our findings, we identify the following opportunity areas for tool builders making qualitative data analysis tools: 1) interactivity, 2) data exploration, and 3) qualitative coding.
% in future qualitative data analysis tools.
% \subsubsection{Interactivity throughout the research process} 
% Our findings identify specific opportunities for designers of tools to aid qualitative research: 
Based on our findings from qualitative researchers interested in engaging with LLMs, tool builders could emphasize 1) interactivity, 2) data exploration, and 3) support for qualitative coding in future qualitative research tools.
Participants consistently mentioned the useful nature of the interactivity LLMs afford through a chat-based interface: they could ideate, obtain feedback, and potentially make requests for data retrieval or formatting in natural language (e.g. \cite{izacard2021leveraging, gao2023retrieval}).
% Some participants mention that the chat interaction helps them \textbf{brainstorm} ideas, \textbf{gut check} findings or intuitions about their research, or  \textbf{see things another way}. 
% Others hope to make data retrieval or formatting requests in natural language through an LLM, which may be increasingly possible as retrieval-augmented generation (RAG) systems improve in fidelity to source materials. \cite{izacard2021leveraging, gao2023retrieval} 
There is a gap in tools that facilitate these use cases in addition to other, more structured uses.
% The risks of designing poorly for interactivity could include several known risks of AI interaction, including risks of anthropomorphizing or overreliant on it, decreasing researcher agency, or using it to decrease engagement with real collaborators.
% \subsubsection{Data exploration}
% Second, researchers also perceive an opportunity to use LLMs to explore their data or see it in a new way. 
% Risks of implementing designs for this use case include presenting just one hegemonic perspective with the tool, but designed thoughtfully, 
Second, LLMs could help researchers see their data in a new way, ideally widening rather than homogenizing the nature of their insights. 
% \subsubsection{Data annotation}
Lastly, qualitative researchers are still interested in using LLMs to conscientiously assist in the coding process.
% , particularly if they have collected interviews themselves and thus already have a deep understanding of it. 
% Making deductive coding tools more readily available for qualitative researchers 
% to facilitate coding for those who may not be able to engineer their own annotation pipeline 
% is an opportunity area.
% Tools could also help scaffold the qualitative coding process for beginners, but some worry that tools may diminish \textit{"essential friction"} (P15) as one learns coding. 
% Tools focused on this use case can be developed to minimize overreliance and align with qualitative research values.
% Tools should help researchers check if the LLM is implementing their desired definitions and codes correctly.
% Across all opportunity areas, LLM-based tools also present risks which must be navigated carefully, such as the risk of anthropomorphizing AI, becoming overreliant on it, decreasing researcher agency, reducing novelty, or the risk of broader misalignment with qualitative research values.
% Next, we highli
% start of section on design principles
% \subsection{Design principles for qualitative research tools incorporating LLMs}
% Researchers want to use tools that align with their values. 
% Some implementations of LLM-based tools can empower researchers, while others may risk dispossessing them or their participants. 
Below are principles for tool designers when considering how to create tools that responsibly address these opportunity areas.

\subsubsection{Design for participant privacy}
Proprietary LLMs have sparked concerns regarding participant privacy. Tools that incorporate LLMs should be explicit about if, when, and how they call external APIs, and whether LLMs used are proprietary or open source. They should also explain to users before data is uploaded how they handle privacy, how they expect users to deanonymize data before using tools, how they scrub or plan to delete data, and whether they use inputs for training models. 
Second, some researchers mention open source models as a potential solution to privacy concerns, but many are unable to implement solutions to this problem themselves. 
As such, there is an opportunity area to support researchers in having greater control and transparency over data privacy. 
More tools that make it easy to locally host, steer, and fine-tune open source models for qualitative research functions could empower researchers.
% in appropriate tasks.
% qualitative researchers feel more confident about ethically using LLMs in qualitative research. 

\subsubsection{Design for intentional use}
The lines between structured tools built for a single task and open-ended chat-based tools are blurring.
% eginning to blur. 
Chat-based LLMs allow  flexibility and interactivity, but do not currently support intentional use for specific tasks in the research process, making slippage into unintentional uses of LLMs easy. Researchers want to maintain agency over if, when, and how they are assisted or influenced by AI tools in their research process. 
% Scaffolding user interaction can help them opt in or out of particular use cases of LLM-based tools, as well as guard against overreliance and inappropriate use.
Tools could be designed to guide users to intentionally select appropriate uses of LLM-based tools, and also offer features that allow researchers agency to develop their own ideas before opting into LLM assistance in any task.

% One way to design for agency is to give researchers ways of opting in or out of AI-based features within platforms, facilitating researchers' own development of theory from the data before providing assistance.
% Designing for agency also involves designing against overreliance on or anthropomorphism of these technologies. 

\subsubsection{Design for transparency and validation}

Tools should give researchers ways to transparently evaluate performance.
% Giving researchers methods of transparent evaluation of the tool can make researchers feel empowered and confident in their decisions throughout the research process if they select the tool for their desired task. 
Appropriate validation varies depending on research values, and on the specific research task.
% and looks different depending on the \textit{task} being accomplished in the work.
For brainstorming, designing for validation could involve designing features to understand or probe outcomes and suggestions.
For positivist qualitative coding tasks, features designing for reproducibility of outcomes, interrater reliability, and error analysis from LLM annotation compared to human coding may be useful.
Features that require researchers to review outputs may help accomplish this goal, depending on the task.

\subsubsection{Design for researcher context} 
Researchers often wish to bring in their own context when performing qualitative analysis, including their past experiences, the texts that influence their work, and the theories in which they ground their work. 
% Tools for qualitative researchers should be designed assuming researchers have unique perspective to contribute to the data. 
Tools could embed unique researcher perspectives with models that are trained or fine-tuned 
% Models could be steered or fine-tuned 
according to particular data sets or perspectives of interest to the researcher.

\subsubsection{Design for deep engagement with data}
Qualitative researchers worry that LLM-based tools could distance them from data rather than deepen their relationship with it, but tools can be explicitly designed to foster `closeness' with data~\cite{gilbert2002going}.
% to elicit deeper engagement with the data.
Many qualitative researchers expressed interest in using LLMs to see their data through another lens, but found that insights and themes produced on their data are generic or repetitive. 
Tools can explore LLMs as a way to produce more unique ways of understanding data, drawing researchers' attention to \textit{less} predictable features of the data, as suggested by~\citet{jiang2021supporting}, or helping researchers examine or challenge their existing theories with direct evidence from the data.
These features may even deepen researchers' understanding of the data.

\subsubsection{Design to consider participant perspectives and interests} 
Researchers are concerned about the variable performance LLMs have across contexts, knowledge domains, cultures, and languages. Tools should work to facilitate better alignment between LLMs and diverse subject populations. When possible, tool builders could give the option to use models that better reflect a target, like models that specialize in a particular language or have specialized training data, and facilitate investigating where there might be errors or bias.
% , in alignment with the recommendation to design for intentionality, agency, and validation.

% \subsection{Limitations [RENAME]}nv
% Nonetheless, our sample leans mixed
\section{Limitations}
While we recruited participants across disciplines, our participant group does not cover all perspectives on qualitative research and its values. 
Disciplines that are even more situated than those represented in our participant group, such as ethnography, may be at greater risk of diminishment by LLM involvement, and those perspectives may not be represented by the voices highlighted here.
% We believe the more contextually situated, slower, and interpretivist qualitative epistemologies are, the more LLMs are likely to be at odds with or threaten their research values.
% We also recruited participants regardless of their level of engagement with LLMs as users or as researchers. 
% This elicited important findings about current use by non-experts, important intuitions from qualitative researchers regardless of their level of use, and insights from experts on responsible AI who avoid LLMs.
% However, some of these researchers, by virtue of not being experts on LLMs and their potential flaws, may have presented a different set of concerns than experts, particularly on the nature of risks from LLMs or their potential capabilities under different conditions of use.
% (Maybe a good place to say we had several researchers who study AI bias, who did present strong concerns).
The fact that our participants were mostly PhD students also likely shapes the findings, since younger people have more positive views of AI~\cite{stein2024attitudes}, and senior researchers are more likely to have ethical concerns about AI research use~\cite{liao2024llms}.
While our participants represent a diverse group of people studying various populations, there was a heavy bias for American universities, which may miss key perspectives from non-Western institutions.
% the set of concerns elicited, level of engagement with these tools, and interest in future use of LLMs in research. 
% We also note our participants largely did not surface experience or opinions on simulated research subjects or automated interviewing, which are garnering significant attention in other research circles.
Despite these limitations, our study highlights an important population wrestling with incorporating transformative technologies into their research processes for the first time.

\section{Conclusion}

LLMs are poised to continue proliferating and permeating  research. 
While LLMs present potential opportunities for assistance in some research tasks, they also surface potential tensions with the values of qualitative research. 
In this climate, we document a sociotechnical breakdown: the functionality of \textit{tooling} is insufficient to reliably preserve participant privacy, the \textit{guidelines and norms} lag behind, and the \textit{social context} increasingly incentivizes the broad use of AI across surfaces.
% As they do so, researchers have agency to decide if, when, and how they will engage with LLMs in qualitative work according to their research values and participants' interests.
Active, intentional consideration from researchers on issues like privacy and model bias in advance of task selection, coupled with thoughtful tool design, can help to maintain agency as researchers decide appropriate uses for LLMs.
% , and thoughtful tool design can create an environment of appropriate, intentional use.
% ChatGPT spurred new ideas for researchers regarding how their research processes might be augmented or threatened, but it should not serve as either a model for what is possible for LLMs in research tooling. 
We synthesized current uses, tensions, and ethical challenges into a set of choices for the intentional incorporation of LLMs in qualitative research.
% and design principles for tooling in support of qualitative researchers. 
We hope that this paper will empower qualitative researchers to leverage LLMs confidently, and even creatively, for their work if they choose.
% These considerations and design principles can help qualitative researchers better navigate choice points as they work to ethically engage with LLMs in their work.

\begin{acks}
    We thank James Eschrich for their formative contributions to the scope, literature, and analysis in this piece. We also thank Mor Naaman, Emily Tseng, Louise Barkhuus, and students and staff at the MIT Center for Constructive Communication for their perspectives on this work. We thank the participants of the ``LLMs as Research Tools" workshop at CHI 2024 for their contributions that informed the direction of this work. 
We acknowledge that we used ChatGPT and Elicit to find specific works of related literature, which we then verified from source documents.
\end{acks}

% \section{Analysis plan}

% The interview data will be annotated by the team of PhD researchers on the team according to thematic categories that emerge from saturation from the interview process, review of data, and collaborative discussion with collaborators. Survey data would be descriptively presented according to the main findings of the qualitative research and goals of the paper. 

% \section{Deliverables}

% The goal will be to run the interviews and survey, and analyze the data in preparation for a submission on our results to the deadline for CHI 2025 in September, 2024.

\bibliographystyle{ACM-Reference-Format}
\bibliography{ref2}

\section{Appendix}

\subsection{Interview Ethics}
We took several ethical concerns into consideration when conducting the interviews. 
All transcriptions were done through Zoom or Adobe Premiere Pro and edited manually by the research team to ensure fidelity to the speaker's words. Both Zoom and Adobe Premiere Pro encrypt user data and delete transcriptions.
Direct participant quotes were never entered into ChatGPT or any other generative AI tool.
Participants were permitted to use video or audio only for the interviews.
Participants were compensated \$15 for each interview.
During each interview, we obtained verbal consent to record and participants could stop the interview or refuse to answer any question.
We informed participants that their anonymity would be preserved, not only by protecting their personally identifiable information, but also by obscuring or altering any highly specific research studies they mentioned.

\subsection{Interview Protocol}
\textit{Section 1: Situating the researcher}
\begin{itemize}
  \item Can you give me a high-level overview of your work?
  \item Can you describe your relationship with qualitative and human-centered research?
  \begin{itemize}
  \item How would you define qualitative research?
  \item How would you define human-centered research?
\end{itemize}
  \item What field are you in or where do you publish?
  \item What would you consider your relationship with Large Language Models to be?
\end{itemize}

\textit{Section 2: LLMs in the Research Process}

These questions will each be asked for the steps \textit{Prep and Brainstorming} and \textit{Data Analysis}. For each stage, researchers will be asked "Have you ever used LLMs for this stage of research?" and if the answer is yes, they are asked the following:
\begin{itemize}
  \item What motivated you to use an LLM for that purpose?
  \item How well did the LLM perform and how did you evaluate its performance?
  \item Did you have any ethical concerns with using LLMs in this stage of the research process?
  \begin{itemize}
  \item If yes, how did you try to mitigate those concerns?
\end{itemize}
  \item Would you use LLMs for this purpose again? How would/wouldn’t you change your use of LLMs for this task going forward?
\end{itemize}
If the answer to the initial question was no, the following questions will be asked:
\begin{itemize}
    \item Why do you think you haven’t used LLMs for this stage of research?
    \item What are your main concerns with others using LLMs in this stage of research?
\end{itemize}

\textit{Section 3: Overall thoughts}

Next, we transition to a higher level discussion of thoughts on LLMs in research.
\begin{itemize}
  \item Overall, what do you think are the promising use cases for using LLMs in interview research? 
  \item On the other side, if anything, what do you fear might be lost from incorporating LLMs into qualitative research? What are your main concerns?
  \item Generally speaking, what role do you perceive the LLM as occupying? Annotator? Co-researcher? Tool?
\end{itemize}

\textit{Section 4: Wrap up}

Finally, we wrap up.
\begin{itemize}
  \item Is there anything else that you’d like me to know or you think is interesting about your usage of LLMs for interview research? 
  \item Do you have any questions for me?
  \item Is there anyone else in your circle who is a qualitative researcher that you might be willing to put me in touch with? If yes, I can send you a follow up email to contact.
\end{itemize}

\end{document}